\documentclass[prd,superscriptaddress,onecolumn,showpacs,showkeys]{revtex4}
\usepackage{eurosym}
\usepackage{amsfonts}
\usepackage{array}
\usepackage{amsthm}
\usepackage{bm}
\usepackage{palatino}
\usepackage[colorinlistoftodos]{todonotes}
\usepackage{mathpazo}
\usepackage{amssymb}
\usepackage{eurosym}
\usepackage{amsmath}
\usepackage{epsfig}
\usepackage{graphics}
\usepackage{color}
\usepackage{graphicx}
\usepackage[colorlinks=true,
            linkcolor=blue,
           urlcolor=blue,
           citecolor=blue]{hyperref}

\def\be{\begin{equation}}
\def\ee{\end{equation}}
\def\beq{\begin{eqnarray}}
\def\eeq{\end{eqnarray}}

\def\B{\mathcal{B}}

\def\bes{\begin{eqnarray}}
\def\ees{\end{eqnarray}}

\begin{document}
\title{Hawking radiation from cubic and quartic black holes via tunneling of GUP corrected scalar and fermion particles}

\author{Wajiha Javed}
\email{wajiha.javed@ue.edu.pk; wajihajaved84@yahoo.com} 
\affiliation{Division of Science and Technology, University of Education, Township-Lahore, Pakistan}

\author{Rimsha Babar}
\email{rimsha.babar10@gmail.com} 
\affiliation{Division of Science and Technology, University of Education, Township-Lahore, Pakistan}

\author{Ali \"{O}vg\"{u}n}
\email{ali.ovgun@pucv.cl}
\homepage[]{https://www.aovgun.com}
\affiliation{Instituto de F\'{\i}sica, Pontificia Universidad Cat\'olica de Valpara\'{\i}%
so, Casilla 4950, Valpara\'{\i}so, Chile}
\affiliation{Physics Department, Faculty of Arts and Sciences, Eastern Mediterranean
University, Famagusta, North Cyprus, via Mersin 10, Turkey}

\date{\today }

\begin{abstract}
We analyze the effect of the generalized uncertainty (GUP) principle on the Hawking radiation
from the hairy black hole in U(1) gauge-invariant scalar-vector-tensor theory by utilizing the semiclassical
Hamilton-Jacobi method. To do so, we evaluate the tunneling probabilities and Hawking
temperature for scalar and fermion particles for the given spacetime of the black holes with cubic and quartic interactions.
For this purpose, we utilize the modified Klein-Gordon equation for the Boson particles and then Dirac equations for the fermion particles, respectively. Next, we examine that the Hawking temperature of the black holes do not depend on the properties of tunneling
particles. Moreover, we present the corrected Hawking temperature of scalar and fermion particles looks similar in both interactions,
but there are different mass and momentum relationships for scalar and fermion particles in
cubic and quartic interactions.
\end{abstract}

\keywords{ Effective Hawking temperature, Modified Klein-Gordon, Modified Dirac equation.}
\pacs{04.20.Jb, 04.62.+v, 04.70.Dy }
\maketitle

\section{Introduction}
As indicated by Einstein's theory of general relativity,
black holes (BHs) are very intense transcendental objects
that twist spacetime so unequivocally that no light or matter
can get out of their grasp. According to NASA, some primordial BHs have
shaped before long after the Big Bang and may be of the estimate of a
single molecule, anyhow, as enormous as a mountain.
In $1960's$, John Wheeler and his
collaborators propounded that BHs "have no hair",
an allegory meaning that BHs trim of all
complex features. In general relativity, no-hair
theorem postulates that BHs are
surprisingly simple objects,
they can only describe by three
exceptional classical parameters for them: their
mass, electromagnetic
charge and angular momentum \cite{A1}-\cite{A5}. The Wheeler’s
hypothesis of no hair theorem was
proven wrong in  Einstein-Yang-Mills theory \cite{A6}.
It is familiar that BHs have hair within the existence of Yang-Mills
fields \cite{A7}-\cite{A9}.

The modification of the general relativity may include scalar fields. In these theories no-hair
theorem still exists. The primary no-scalar-hair theorem has connected to
the massless scalar \cite{A10} and to the impartial scalar fields with a monotonous increasing
self-inducting potential \cite{A11}.
Heisenberg and Sujikawa \cite{A18} investigated the
solution of hairy BH under U(1) gauge-invariant
scalar-vector-tensor (SVT) theories for cubic and quartic order
scalar-vector interactions. Herdeiro and Radu \cite{A19} have
introduced a new family of BHs
with scalar hair, that's persistently associated to the Kerr
family and gave a subjectively new case of hairy BHs entitled
Kerr BHs with scalar hair.
It is familiar that BHs can grow
scalar hair in the presence of matter in their
proximity \cite{f2,f3}, complex scalar with time dependent stages \cite{f4}-\cite{f6},
or on the off chance that when the asymptotics are cosmological or anti-de Sitter
\cite{f7}-\cite{f9}.

General relativity extensions to generalized gravitation theories
extensively presents new parametric quantities apart from two tensor
polarization \cite{f10}. The development of SVT theories with Horndeski
theories was also carried out for both the U(1) gauge and
non-gauge invariant cases \cite{f15}. The new parametric quantities emerging in SVT
theories are important for the
thermodynamical properties of BHs and expansion of universe.

Stephen Hawking ($1970$), presented his idea about black body
radiations, which is known as Hawking radiation. All BHs losses their mass as
apparitional quantum particles \cite{Hawking:1974rv,Hawking}.
Due to the continuous process of Hawking radiation a BH evaporates \cite{Tia:2016,Unruh:1976db}.
Hawking evaluated the BH physics in a curved spacetime, under quantum field theory depending upon
the Heisenberg uncertainty principle (GUP). The Hawking radiation through quantum tunneling
strategy of the radiated particles
from a BH has been investigated \cite{A12}-\cite{A15}. Moreover, the
Hawking radiation through tunneling
phenomenon for numerous BHs
has been investigated, abundantly, in $(2+1)$ and $(3+1)$-dimensional \cite{A16,A17,Parikh:1999mf,
Akhmedov:2006pg,Kerner:2006vu,Angheben:2005rm,
Singleton:2011vh,Sakalli:2018nug,Gonzalez:2017zdz,Jusufi:2017trn,Kuang:2017sqa,Ovgun:2017iyb,
Jusufi:2017vhz,Sakalli:2017ewb,Jusufi:2016hcn,Ovgun:2016roz,Sakalli:2016cbo,Sakalli:2016mnk,
Sakalli:2016ojo,Jusufi:2015mii,Ovgun:2015box,Ovgun:2015jna,Sakalli:2015jaa,Sakalli:2015mka,
Sakalli:2015taa,Sakalli:2015nza,Sakalli:2014sea,Sakalli:2014bfa,mali,mali2,mali3,mali4,
Pasaoglu:2009te,Sakalli:2010yy,Sakalli:2012zy,Sakalli:2012xd,Mazharimousavi:2009nc,
Meitei:2018mgo,Rodrigue:2018dlg,Ma:2018xpw,Ma:2017odv,Sefiedgar:2017ybf,Gecim:2017nbh,
Li:2017kze,Feng:2017lkn,Singh:2017car,Haldar:2017viz,Gecim:2017vgx,Li:2017fss,Li:2017pyf,
Gecim:2017lxx,Javed:2017saf,Rizwan:2016ldf,Saleh:2015tcp,Ding:2011zzb}.
Many authors have studied quantum tunneling approach for particles
with different spins such as vectors (bosons), scalar and fermions
through the horizons of various BHs, wormholes and  other celestial
objects, they also calculated their corresponding Hawking
temperatures.

The GUP plays very important role
to evaluate the quantum gravity effects (quantum corrections). To
acquire the GUP effects on Hawking temperature, we will utilize
corrections of Klein-Gordan and Dirac equations by considering
quantum effects.
Black holes are the main experimentation field to investigate the quantum
gravity effects and lots of literature on BH thermodynamical properties to study
the quantum gravity effects under GUP. The BH thermodynamics have
moreover been explored within the system of GUP \cite{B45}-\cite{B51}. Nozari and
Mehdipour \cite{B52} have investigated the tunneling phenomenon
under GUP effects for Schwarzschild BH and also evaluated its modified tunneling rate.
\"{O}vg\"{u}n et al. \cite{B55}
have calculated the tunneling of massive
spin-1 and spin-0 particles for a warped Dvali-Gabadadze-Porrati
(DGP) gravity BH and also discussed the effects of GUP for both type
of particles. \"{O}vg\"{u}n, Javed and Ali \cite{B56} have found the
tunneling rate of charged massive bosons for various types of BHs
surrounded by the perfect fluid in Rastall theory. Sharif
and Javed \cite{a4} have considered tunneling phenomenon for fermion particles
through the event horizons for the study of Hawking
temperature. They also discussed the Hawking radiation
through tunneling phenomenon for fermion particles from a pair of charged
accelerating and rotating BH around NUT parameter \cite{a2}. Furthermore, they
analyzed the corresponding Hawking temperature of these BHs. Sharif and Javed
\cite{a7} have also investigated the quantum corrections for regular
BHs, i.e., Bardeen and ABGB.

In the continuation of the previous work, we will investigate the
quantum corrections of massive scalar and fermion particles by using
GUP effects. A brief outline of paper is given as follows:
In Section \textbf{II}, we provide the
detail of hairy BH in the existence of U(1) gauge scalar-vector-tensor theories.
In Section \textbf{III}, we explore the GUP-corrected Klein-Gordan equation for
cubic and quartic interactions and examine the
quantum corrections of Hawking temperature for massive
scalar particles for BH solution surrounded by SVT theory.
Whereas, Section \textbf{IV} is devoted to the same analysis with
Dirac equation for fermion particles. Finally, the results of this paper are
summarized in Section \textbf{V}.

\section{Hairy Black Hole under Gauge-invariant Scalar-vector-tensor Theory}

Black holes play a very essential role in general relativity. The Gibbons solution which depicts a Reissner-Nordstorm BH
with a non-trivial expansion field is considered as the first hairy BHs.
Whereas numerous ways run out scalar field hair in asymptotically flat spacetime \cite{f16}. However, more hairy BHs
have gotten in models motivated by string theory and counting expansion, curvature
correction and many other. By considering positive or negative, cosmological constant
asymptotically (anti)-de-Sitter hairy BHs are obtained. Hairy BHs generally occur in natural models.
However, large hairy BHs are regularly unstable and during perturbation they loose their hair,
although the stable one are exceptionally little. The hairy BHs yields
as cosmologically large,
which conflicts the observation \cite{f17}.
It will be very interesting to evaluate the tunneling probability and corrected Hawking temperature from such a BH.
The metric of hairy BH for cubic and quartic interactions is defined as \cite{A18}
\begin{equation}
ds^2=-Z(r)dt^2+{h^{-1}(r)}dr^2+r^{2}(d\theta^{2}+sin^{2}\theta d\phi^{2}),\label{W01}
\end{equation}
where for cubic interaction
\begin{eqnarray*}
Z(r)=1-\frac{2M}{r}+\frac{Q^{2}}{2M^{2}_{pl}r^{2}}+\frac{3\beta^{2}_{3}Q^{4}}
{14M^{2}_{pl}r^{8}}+O\left(\frac{1}{r^{9}}\right),\\
h(r)=1-\frac{2M}{r}+\frac{Q^{2}}{2M^{2}_{pl}r^{2}}-\frac{2\beta^{2}_{3}Q^{4}}
{7M^{2}_{pl}r^{8}}+O\left(\frac{1}{r^{9}}\right).
\end{eqnarray*}
For quartic interaction
\begin{eqnarray}
Z(r)&=&1-\frac{2M}{r}+\frac{Q^{2}}{2M^{2}_{pl}r^{2}}-\frac{2\beta_{4}Q^{2}}
{M^{2}_{pl}r^{4}}+\frac{2\beta_{4}MQ^{2}}{M^{2}_{pl}r^{5}}-\frac{3\beta_{4}Q^{4}}
{5M^{4}_{pl}r^{6}}+\frac{256\beta^{2}_{4}MQ^{2}}{7M^{2}_{pl}r^{7}}\nonumber\\
&+&\frac{3Q^{2}(M^{2}_{pl}Q^{2}\beta^{2}_{3}-28\beta^{2}_{4}Q^{2}-256\beta^{2}_{4}M^{2}M^{2}_{pl})}{14M^{4}_{pl}r^{8}},\nonumber\\
h(r)&=&1-\frac{2M}{r}+\frac{Q^{2}}{2M^{2}_{pl}r^{2}}-\frac{2\beta_{4}MQ^{2}}
{M^{2}_{pl}r^{5}}+\frac{2\beta_{4}Q^{4}}{5M^{4}_{pl}r^{6}}-\frac{2Q^{2}(\beta^{2}_{3}Q^{2}-64\beta^{2}_{4}M^{2})}
{7M^{2}_{pl}r^{8}}.\label{g1}
\end{eqnarray}
Here, $M$ stands for BH mass and $Q$ represents the charge of BH.

\section{Scalar Particle tunneling via Modified Klein-Gordan Equation}

This section is based on the effects of GUP on
the tunnling of massive scalar particles from the
hairy BH solutions for cubic and quartic interactions.

\subsection{Cubic Interaction for Scalar Particles}

This section is devoted to study the tunneling phenomenon of scalar particles
for given BH with cubic interaction
and analyze the corresponding corrected Hawking temperature.
For this purpose, the modified Klein-Gordan equation is given by \cite{Ganim:2017}
\begin{equation}
-(\iota\breve{\hbar})^{2}\partial^{t}\partial_{t}\Phi=[(-\iota\breve{\hbar})^{2}
\partial^{i}\partial_{i}+m_{0}^{2}][1-2\breve{\beta}(-\iota\breve{\hbar})^{2}
\partial^{i}\partial_{i}+m_{0}^{2}]\Breve{\Phi},\label{A3}
\end{equation}
The wave function $\Breve{\Phi}$ for scalar field is given as
\begin{equation}
\Breve{\Phi}(t,r,\theta,\varphi)=\left[\frac{\imath}{\breve{\hbar}}I(t,r,\theta,\varphi)\right],
\end{equation}
we consider just first order term in $\breve{\hbar}$, so the above
Eq.(\ref{A3}) becomes
\begin{eqnarray}
&&\frac{1}{Z(r)}(\partial_{t}I)^{2}=\left[h(r)
(\partial_{r}I)^{2}+\frac{1}{r^{2}}(\partial_{\theta}I)^{2}+
\frac{1}{r^{2}sin^{2}\theta}(\partial_{\varphi}I)^{2}+m_{0}^{2}\right]\times\nonumber\\
&&\left[1-2\breve{\beta}\left(h(r)(\partial_{r}I)^{2}+\frac{1}{r^{2}}
(\partial_{\theta}I)^{2}+\frac{1}{r^{2}sin^{2}\theta}(\partial_{\varphi}I)^{2}+m_{0}^{2}\right)\right].\label{R4}
\end{eqnarray}
To solve this equation, the particle's action is defined as
\begin{equation}
I=-Et+W(r,\theta)+j\varphi.
\end{equation}
Here, $W(r,\theta)$ cannot be parted
as $W(r)\Theta(\theta)$. For sake of simplicity, we can fix
the angle $\theta$ at a particular value of $\theta_{0}$.

At $\theta=\theta_{0}$, Eq.(\ref{R4}) gets the following form
\begin{equation}
\breve{A}(\partial_{r}W)^{4}+\breve{B}(\partial_{r}W)^{2}+\breve{C}=0,\label{R5}
\end{equation}
here
\begin{eqnarray*}
\breve{A}&=&-2\breve{\beta} h^{2}(r),~~~\breve{B}=h^{2}(r)\left(1-
\frac{4\breve{\beta} j^{2}}{r^{2}sin^{2}\theta}-4\breve{\beta} m_{0}^{2}\right),\\
\breve{C}&=&m_{0}^{2}+\frac{j^{2}}{r^{2}sin^{2}\theta}-\frac{2\breve{\beta} j^{4}}{r^{4}sin^{4}\theta}-
 \frac{4\breve{\beta} m_{0}^{2}j^{2}}{r^{2}sin^{2}\theta}-2\breve{\beta}
m_{0}^{4}-\frac{E^{2}}{Z(r)}.
\end{eqnarray*}
After solving Eq.(\ref{R5}), we get
\begin{eqnarray}
W_{\pm}(r)&=&\pm\int\frac{dr}{\sqrt{Z(r)}}\left[1+\breve{\beta}\left(m_{0}^{2}+\frac{E^{2}}{Z(r)}
+\frac{j^{2}}{r^{2}sin^{2}\theta}\right)\right]\nonumber\\
&\times&
\sqrt{E^{2}-Z(r)\left(m_{0}^{2}-\frac{j^{2}}{r^{2}sin^{2}\theta}\right)
-2\breve{\beta} Z(r)\left(\frac{j^{4}}{r^{4}sin^{4}\theta}+
2m_{0}^{2}Z(r)\frac{j^{2}}{r^{2}sin^{2}\theta}+m^{4}\right)}.
\end{eqnarray}
After ignoring higher order terms of
$\breve{\beta}$, we solve above integral to calculate the imaginary part of function at event horizon
\begin{equation}
\textit{Im}W(r_{+})=\pm\pi
\left(\frac{Er_{+}^{2}}{\Delta,_{r}(r_{+})}\right)(1+\breve{\beta}\Xi),
\end{equation}
where
\begin{equation*}
\Xi=\left(m^{2}+\frac{E^{2}}{F^{'}(r_{+})}+\frac{j^{2}csc^{2}\theta}{r^{2}_{+}}\right),
\end{equation*}
Here, $W_{+}$ and $W_{-}$ are the radial functions for the
outgoing and incoming particles, respectively. The tunneling probability for
scalar particles at $r=r_{+}$ is given as follows
\begin{eqnarray}
\Breve{\Gamma}&=&\frac{\Breve{\Gamma}_{(out)}}{\Breve{\Gamma}_{(in)}}=
\frac{\exp\left[-\frac{2}{\breve{\hbar}}(ImW_{+})\right]}
{\exp\left[-\frac{2}{\breve{\hbar}}(ImW_{-})\right]}=
\exp\left[-\frac{4}{\breve{\hbar}}ImW_{+}\right],\nonumber\\
&=&\exp\left[-\frac{4\pi
r^{2}_{+}}{\breve{\hbar}\Delta,_{r}(r_{+})}(E)\times(1+\breve{\beta}\Xi)\right].
\end{eqnarray}
For $\breve{\hbar}=1$ and utilizing Boltzmann factor $\Breve{\Gamma}=
\exp\left(\frac{E}{T^{'}_{H}}\right)$, the modified
temperature can be obtained as
\begin{equation}
T^{'}_{H}=\frac{\Delta,_{r}(r_{+})}{4\pi
r^{2}_{+}(1+\breve{\beta}\Xi)}=T_{H}(1-\breve{\beta}\Xi),\label{B8}
\end{equation}
where
\begin{equation*}
T_{H}=\frac{1}{4\pi r^{2}_{+}}\left[\frac{2M}{r^{2}_{+}}-\frac{Q^{2}}{M^{2}_{pl}r^{3}_{+}}
-\frac{12 Q^{4}\beta^{2}_{3}}{7M^{2}_{pl}{r^{9}_{+}}}-O\left(\frac{1}{r^{10}_{+}}\right)\right],
\end{equation*}
which is the original Hawking temperature of a corresponding BH.

\subsection{Quartic Interaction for Scalar Particles-I}

Following the same procedure given in the preceding subsection \textbf{A}, for this line element (\ref{g1}), we can calculate
the corrected Hawking temperature with the effects of quantum gravity for this BH.
The modified Hawking temperature for quartic interactions can be deduced as
\begin{equation}
T^{'}_{H}=\frac{\Delta,_{r}(r_{+})}{4\pi
r^{2}_{+}(1+\breve{\beta}\Xi)}=T_{H}(1-\breve{\beta}\Xi),\label{B9}
\end{equation}
where the original Hawking temperature is
\begin{eqnarray*}
T_{H}&=&\frac{1}{4\pi r^{2}_{+}}\left[\frac{2M}{r^{2}_{+}}-\frac{Q^{2}}{M^{2}_{pl}r^{2}_{+}}
-\frac{256M Q^{2}\beta^{2}_{4}}{M^{2}_{pl}{r^{8}_{+}}}-\frac{10 \beta_{4} M Q^{2}}
{M^{2}_{pl}r^{6}}+\frac{8\beta_{4}Q^{2}}{M^{2}_{pl}r^{5}}+
\frac{18\beta_{4}Q^{4}}{5M^{4}_{pl}r^{7}}\right.\\
&-&\left.\frac{12Q^{2}(M^{2}_{pl}Q^{2}\beta^{2}_{3}-28\beta^{2}_{4}Q^{2}
-256\beta^{2}_{4}M^{2}M^{2}_{pl})}{7M^{4}_{pl}r^{9}}\right].
\end{eqnarray*}

\subsection{Quartic Interaction for Scalar Particles-II}

The line element of BH for Quartic interactions is defined as
\begin{equation}
ds^2=-Z(r)dt^2+{h^{-1}(r)}dr^2+r^{2}(d\theta^{2}+sin^{2}\theta d\breve{\Phi}^{2}), \label{g3}
\end{equation}
where
\begin{eqnarray*}
Z(r)&=&(1-\mu)\left(\frac{r}{r_{+}}-1\right)-\frac{1-2\mu+12\beta_3^2\mu^2(1-\mu)+
8\bar{\beta}_4\mu(1-2\mu)}{1+8\bar{\beta}_4}\left(\frac{r}{r_{+}}-1\right)^2\\
h(r)&=&(1-\mu)\left(\frac{r}{r_{+}}-1\right)-\frac{1-2\mu-4\beta_3^2\mu^2(1-\mu)+
8\bar{\beta}_4\mu(1-2\mu)}{1+8\bar{\beta}_4}\left(\frac{r}{r_{+}}-1\right)^2.
\end{eqnarray*}
where $\tilde{\beta}_{3}=\frac{\beta_{3}M_{pl}^{2}}{r_+^2}$ and $\bar{\beta}_{4}=\frac{\beta_{4}M_{pl}^{2}}{r_+^4}$.

Using the same formalism as defined earlier in (\textbf{A}) for this line element, we obtain
the corrected Hawking temperature with the effect of quantum gravity for this BH.
The modified Hawking temperature for quartic interactions is deduced as
\begin{equation}
T^{'}_{H}=\frac{\Delta,_{r}(r_{+})}{4\pi
r^{2}_{+}(1+\breve{\beta}\Xi)}=T_{H}(1-\breve{\beta}\Xi),\label{n9}
\end{equation}
where the original Hawking temperature is
\begin{eqnarray*}
T_{H}&=&\frac{1}{r_+^3(r_+^4+8M_{pl}^2\beta_4)^2}\left[16\mu M_{pl}^2r_{+}\left(-3(\mu-1)\mu M_{pl}^2\beta_3^2\right.\right.\\
&+&\left.2(1-2\mu)\beta_4\right)(r-r_+)^2
(r_+^4+8M_{pl}^2\beta_4)-32M_{pl}^2\beta_4r_+^4(1-2\mu)(r-r_+)\\
&-&4\mu M_{pl}^2r_+\left(-3(\mu-1)\mu M_{pl}^2\beta_3^2
+2(1-2\mu)\beta_4\right)-r_+(\mu-1)(r-r_+)\\
&\times&(r_+^4+8M_{pl}^2\beta_4)^2-2r_+^2(r_+^4+8M_{pl}^2\beta_4)(1-2\mu)-4\mu M_{pl}^2(r-r_+)\\
&\times&\left.\left(-3(\mu-1)\mu M_{pl}^2\beta_3^2+2(1-2\mu)\beta_4\right)\right].
\end{eqnarray*}
\section{The Fermion Particles Tunneling via Modified Dirac Equation}

In this section, we will foccus on studying the effects of GUP on the
tunneling procedure of fermion particles from the hairy BH for cubic
and quartic interactions.

\subsection{Cubic Interactions for Fermion Particles}

In order to investigate the quantum tunneling of fermion particles for hairy BH,
the Dirac equation is given as follows \cite{Gecim:2017nbh}
\begin{equation}
i\tilde{\gamma}^{\nu}\left(\partial_{\nu}+\tilde{\Omega}_{\nu}\right)\Psi
+\frac{m_0}{\hbar}\Psi=0,~~~\nu=0,1,2,3\label{C1}
\end{equation}
where
\begin{eqnarray}
\tilde{\Omega}_{\nu}=
\frac{i}{2}\mathcal{\tilde{\omega}}_{\nu}^{~~\alpha\beta}\tilde{\Sigma}_{\alpha\beta},
~~~\tilde{\Sigma}_{\alpha\beta}=\frac{1}{4}i[\tilde{\gamma}^{\alpha},
\tilde{\gamma}^{\beta}],~~~\{\tilde{\gamma}^{\alpha},\tilde{\gamma}^{\beta}\}=2\tilde{\eta}^{\alpha\beta},
\end{eqnarray}
here $\mathcal{\tilde{\omega}}_{\nu}^{~~\alpha\beta}$ represents the
spin connection which can be defined by the following relation of tetrad $e^{\lambda}_{~~\beta}$, i.e.,
\begin{equation}
\mathcal{\tilde{\omega}}_{\nu~~\beta}^{~\alpha}=e_{\mu}^{~~\alpha}
e^{\lambda}_{~~\beta}\tilde{\Gamma}^{\mu}_{\nu\lambda}-
e^{\lambda}_{~~\beta}\partial_{\nu}e_{\lambda}^{~~\alpha},
\end{equation}
and the $\tilde{\gamma}^{\nu}$'s in curved space-time can be defined as
\begin{equation}
\tilde{\gamma}^{\nu}=e^{\nu}_{~~\alpha}\tilde{\gamma}^{\alpha},~~~~~
\{\tilde{\gamma}^{a},\tilde{\gamma}^{b}\}=2\tilde{g}^{ab},
\end{equation}
The modified Dirac equation is defined as
\begin{equation}
-i \tilde{\gamma}^{0}\partial_{0}\tilde{\Psi}=\left(i \tilde{\gamma}^{i}
\partial_{i}+i \tilde{\gamma}^{\mu}\tilde{\Omega}_{\nu}+\frac{m_{0}}{\tilde{\hbar}}\right)
\left(1+\tilde{\beta} \tilde{\hbar}^{2}\partial_{j}\partial^{j}-\tilde{\beta }m_{0}^{2}\right)\tilde{\Psi},
\end{equation}
the above equation can be rewrite in the following form as
\begin{eqnarray}
&&\left[i \tilde{\gamma}^{0}\partial_{0}+i \tilde{\gamma}^{i}(1-
\tilde{\beta} m_{0}^{2})\partial_{i}+i\tilde{\beta}
\tilde{\hbar}^{2} \tilde{\gamma}^{i}\partial_{i}
(\partial_{j}\partial^{j})+\frac{m_{0}}{\tilde{\hbar}}(1+\tilde{\beta}
\tilde{\hbar}^{2}\partial_{j}\partial^{j}-\tilde{\beta} m_{0}^{2})\right.\nonumber\\
&&-\left.i\tilde{\gamma}^{\mu}\Tilde{\Gamma}_{\mu}\left(1+\tilde{\beta}\tilde{\hbar}^{2}
\partial_{j}\partial^{j}-\tilde{\beta} m_{0}^{2}\right)\right]\tilde{\Psi},\label{m1}
\end{eqnarray}
where $\tilde{\Psi}$ is generalized Dirac spinor.
The Dirac gamma  matrices $\gamma^{\nu}$'s can be defined as
\begin{eqnarray}
&&\tilde{\gamma}^{t}=\frac{1}{\sqrt{Z(r)}}\left({\begin{array}
{cc}i & 0\\0 & -i\\ \end{array}}\right),~~~\tilde{\gamma}^{r}
=\sqrt{h(r)}\left({\begin{array}{cc}0 & \tilde{\sigma}^{3}\\\tilde{\sigma}^{3} & 0\\
\end{array}}\right),\nonumber\\
&&\tilde{\gamma}^{\theta}=\sqrt{g^{\theta \theta}}\left({\begin{array}
{cc}0 & \tilde{\sigma}^{1}\\\tilde{\sigma}^{1} & 0\\ \end{array}}\right),~~~~
\tilde{\gamma}^{\phi}=\sqrt{g^{\varphi \varphi}}\left({\begin{array}
{cc}0 & \tilde{\sigma}^{2}\\\tilde{\sigma}^{2} & 0\\ \end{array}}\right),\nonumber
\end{eqnarray}
where $\sqrt{g^{\theta \theta}}=\frac{1}{r}$ \& $\sqrt{g^{\varphi \varphi}}=\frac{1}{r\sin\theta}$
and Pauli sigma matrices $\tilde{\sigma}$'s can be expressed as follows
\begin{equation}
\tilde{\sigma}^{1}=\binom{0~~~~~1}{1~~~~~0},~~~\tilde{\sigma}^{2}=\binom{0~~-i}{i~~~~~0},
~~~\tilde{\sigma}^{3}=\binom{1~~~~0}{0~-1} \label{A3}.\nonumber
\end{equation}
The wave functions can be defined as
\begin{eqnarray}
\tilde{\Psi}(t,r,\theta,\varphi)&=&\left[{\begin{array}{cccc}
A(t,r,\theta,\varphi) \\ 0\\B(t,r,\theta,\varphi) \\ 0\\ \end{array}}\right]
\exp\left[\frac{i}{\hbar}(t,r,\theta,\varphi)\right],\label{c5}
\end{eqnarray}
where $ A(t,r,\theta,\varphi)$ and $ B(t,r,\theta,\varphi)$ are arbitrary functions of
spacetime coordinates.Using Eq.(\ref{c5}) and values of gamma matrices in Eq.(\ref{m1}), we get the following
set of equations for first order in $\tilde{\hbar}$ and $\tilde{\beta}$, i.e.,
\begin{eqnarray}
&&-\frac{i A}{\sqrt{Z(r)}}(\partial_{t}I)-B\sqrt{h(r)}(1-\tilde{\beta} m_0^{2})
(\partial_{r}I)+B\tilde{\beta}\sqrt{h(r)}(\partial_{r}I)\left[h(r)
(\partial_{r}I)^{2}+g^{\theta \theta}(\partial_{\theta}I)^{2}+g^{\varphi \varphi}
(\partial_{\varphi}I)^{2}\right]\nonumber\\
&&-Am_0\tilde{\beta}\left[h(r)(\partial_{r}I)^{2}+g^{\theta \theta}(\partial_{\theta}I)^{2}+
g^{\varphi \varphi}(\partial_{\varphi}I)^{2}\right]+(1-\tilde{\beta} m_0^{2})Am_0=0,\label{C12}\\
&&\frac{i B}{\sqrt{Z(r)}}(\partial_{t}I)-A\sqrt{h(r)}(1-\tilde{\beta} m_0^{2})(\partial_{r}I)+
Bm_0(1-\tilde{\beta} m_0^{2})+\nonumber\\
&&A\beta\sqrt{h(r)}(\partial_{r}I)\left[h(r)(\partial_{r}I)^{2}+
g^{\theta \theta}(\partial_{\theta}I)^{2}+g^{\varphi \varphi}
(\partial_{\varphi}I)^{2}\right]\nonumber\\
&&-m_0B\tilde{\beta}\left[h(r)(\partial_{r}I)^{2}+
g^{\theta \theta}(\partial_{\theta}I)^{2}+g^{\varphi \varphi}
(\partial_{\varphi}I)^{2}\right]=0,\label{C13}\\
&&B\left[-\sqrt{g^{\theta \theta}}(1-\tilde{\beta} m_0^{2})
(\partial_{\theta}I)-i(1-\tilde{\beta} m_0^{2})
\sqrt{g^{\varphi \varphi}}(\partial_{\varphi}I)+\tilde{\beta} (\partial_{\theta}I)
\sqrt{g^{\theta \theta}}\right.\nonumber\\
&&\left.\left\{h(r)(\partial_{r}I)^{2}+g^{\theta \theta}
(\partial_{\theta}I)^{2}+g^{\varphi \varphi}(\partial_{\varphi}I)^{2}\right\}+
i\tilde{\beta} \sqrt{g^{\varphi \varphi}}(\partial_{\varphi}I)\right.\nonumber\\
&&\left.\left\{h(r)(\partial_{r}I)^{2}+g^{\theta \theta}
(\partial_{\theta}I)^{2}+g^{\varphi \varphi}(\partial_{\varphi}I)^{2}\right\}\right]=0,\label{C14}\\
&&A\left[-\sqrt{g^{\theta \theta}}(1-\tilde{\beta} m_0^{2})(\partial_{\theta}I)-
i(1-\tilde{\beta} m_0^{2}) \sqrt{g^{\varphi \varphi}}(\partial_{\varphi}I)
+\tilde{\beta}(\partial_{\theta}I)\sqrt{g^{\theta \theta}}\right.\nonumber\\
&&\left.\left\{h(r)(\partial_{r}I)^{2}+g^{\theta \theta}
(\partial_{\theta}I)^{2}+g^{\varphi \varphi}(\partial_{\varphi}I)^{2}\right\}+
i B\sqrt{g^{\varphi \varphi}}(\partial_{\varphi}I)\right.\nonumber\\
&&\left.\left\{h(r)(\partial_{r}I)^{2}+g^{\theta \theta}
(\partial_{\theta}I)^{2}+g^{\varphi \varphi}(\partial_{\varphi}I)^{2}\right\}\right]=0.\label{C15}
\end{eqnarray}
By using separation of variables technique, we assume
\begin{equation}
\tilde{I}(t,r,\theta,\varphi)=-Et+W(r)+\Theta(\theta,\phi),
\end{equation}
where $E$ represents energy of the particle,
and $J_{\theta}=\partial_{\theta}\Theta, J_{\phi}=\partial_{\phi}\Theta$.\label{C16}

Using Eq.(\ref{C16}) in Eqs.(\ref{C12})-(\ref{C15}) and by focusing on
Eqs.(\ref{C14}) and (\ref{C15}), we observe that they are
homogeneous after dividing by $A$ and $B$. They can be rewritten as
\begin{eqnarray}
&&\left\{ \tilde{\beta} h_{r}(r_{+})W'^{2}+ g^{\theta \theta} \tilde{\beta}
J_{\theta}^{2}+\tilde{\beta} g^{\varphi \varphi}J_{\varphi}^{2} -(1-\tilde{\beta}
m_0^{2})\right\}\nonumber\\&&\times\left[\sqrt{g^{\theta \theta}}J_{\theta}
+i\sqrt{g^{\varphi \varphi}}J_{\varphi}\right]=0,\label{C17}
\end{eqnarray}
here $W'=\partial_{r}W$, $J_{\theta}=\partial_{\theta}\Theta$
and $\partial_{\varphi}=\partial_{\varphi}\Theta$.

In Eq.(\ref{C17}), $\tilde{\beta}$ indicates the quantum gravity effects so
it cannot be considered as zero, thus the term in large bracket
equals to zero and provide the solution of $\Theta$. From
Eq.(\ref{C17}), we can write the following expression
\begin{equation}
\left[\sqrt{g^{\theta \theta}}J_{\theta}
+i\sqrt{g^{\varphi \varphi}}J_{\varphi}\right]=0.\label{C18}
\end{equation}
After removing $A$ and $B$ from Eqs.(\ref{C12}) and (\ref{C13}), we
get homogeneous equations, which generate
\begin{equation}
U_{6}(\partial_{r}W)^{6}+U_{4}(\partial_{r}W)^{4}
+U_{2}(\partial_{r}W)^{2}+U_{0}=0,\label{A4}
\end{equation}
where
\begin{eqnarray}
&&U_{6}=\tilde{\beta}^{2}h^{3}Z,\nonumber\\
&&U_{4}=\tilde{\beta} h^{2}Z(m_0^{2}\tilde{\beta}+2\tilde{\beta} S-2),\nonumber\\
&&U_{2}=hZ\left[(1-\tilde{\beta} m_0^{2})^{2}+\tilde{\beta}(2m_0^{2}-2m_0^{4}
\tilde{\beta}-2S+\tilde{\beta} S^{2})\right],\nonumber\\
&&U_{0}=m_0^{2}(1-\tilde{\beta} m_0^{2}-\tilde{\beta} S)^{2}Z-E^{2},\nonumber\\
&&S=g^{\theta \theta}J_{\theta}^{2}+g^{\varphi \varphi}J_{\varphi}^{2}.\nonumber
\end{eqnarray}
From Eq.(\ref{C18}), we note that $S=0$. Considering $\beta$ only
for first order and solving the above Eq.(\ref{A4}) at event horizon, which
gives
\begin{equation}
W(r)=\pm\int\frac{1}{\sqrt{hZ}}\sqrt{m_0^{2}(1-2\tilde{\beta} m_0^{2})Z
+E^{2}}\left[1+\tilde{\beta}\left(m_0^{2}+\frac{E^{2}}{Z}\right)\right]dr.\nonumber
\end{equation}
The above equation implies
\begin{equation}
W(r)=\pm i\pi\frac{Er^2_+}
{\Delta_{r}(r_{+})}(1+\tilde{\beta}\Sigma),
\end{equation}
where
\begin{equation*}
\Sigma=\left(m_0^{2}+\frac{E^{2}}{Z'(r_+)}\right).
\end{equation*}
The tunneling rate of
scalar particles at $r=r_{+}$ can be calculated as
\begin{eqnarray}
\Tilde{\Gamma}&=&\frac{\Tilde{\Gamma}_{(out)}}{\Tilde{\Gamma}_{(in)}}=
\frac{\exp\left[-\frac{2}{\tilde{\hbar}}(ImW_{+})\right]}
{\exp\left[-\frac{2}{\tilde{\hbar}}(ImW_{-})\right]}=
\exp\left[-\frac{4}{\tilde{\hbar}}ImW_{+}\right],\nonumber\\
&=&\exp\left[-\frac{4\pi
r^{2}_{+}}{\tilde{\hbar}\Delta,_{r}(r_{+})}(E)\times(1+\tilde{\beta}\Sigma)\right].
\end{eqnarray}
The modified Hawking
temperature can be obtained as
\begin{equation}
T^{'}_{H}=\frac{\Delta,_{r}(r_{+})}{4\pi
r^{2}_{+}(1+\tilde{\beta}\Sigma)}=T_{H}(1-\tilde{\beta}\Sigma),\label{m8}
\end{equation}
where
\begin{equation*}
T_{H}=\frac{1}{4\pi r^{2}_{+}}\left[\frac{2M}{r^{2}_{+}}-\frac{Q^{2}}{M^{2}_{pl}r^{3}_{+}}
-\frac{12 Q^{4}\beta^{2}_{3}}{7M^{2}_{pl}{r^{9}_{+}}}-O\left(\frac{1}{r^{10}_{+}}\right)\right],
\end{equation*}
which is the original Hawking temperature of a corresponding BH.

\subsection{Quartic Interaction for Fermion Particles-I}

Following the same process, 
for line element (\ref{g1}) of quartic interactions, we calculate the
corrected Hawking temperature under the effect of quantum gravity.
The modified Hawking temperature for quartic interactions of fermion particles is deduced as
\begin{equation}
T^{'}_{H}=\frac{\Delta,_{r}(r_{+})}{4\pi
r^{2}_{+}(1+\tilde{\beta}\Sigma)}=T_{H}(1-\tilde{\beta}\Sigma),\label{m2}
\end{equation}
where
\begin{eqnarray*}
T_{H}&=&\frac{1}{4\pi r^{2}_{+}}\left[\frac{2M}{r^{2}_{+}}-\frac{Q^{2}}{M^{2}_{pl}r^{2}_{+}}
-\frac{256M Q^{2}\beta^{2}_{4}}{M^{2}_{pl}{r^{8}_{+}}}-\frac{10 \beta_{4} M Q^{2}}
{M^{2}_{pl}r^{6}}+\frac{8\beta_{4}Q^{2}}{M^{2}_{pl}r^{5}}+
\frac{18\beta_{4}Q^{4}}{5M^{4}_{pl}r^{7}}\right.\\
&-&\left.\frac{12Q^{2}(M^{2}_{pl}Q^{2}\beta^{2}_{3}-28\beta^{2}_{4}Q^{2}
-256\beta^{2}_{4}M^{2}M^{2}_{pl})}{7M^{4}_{pl}r^{9}}\right],
\end{eqnarray*}
which is the original Hawking temperature of the corresponding BH.

\subsection{Quartic Interaction for Fermion Particles-II}

By following the same procedure, given in the preceding Section (\textbf{A}) for line element (\ref{g3}) of
quartic interactions, we calculate the corrected Hawking temperature under
the effect of quantum gravity.
The modified Hawking temperature for quartic interactions of fermion particles is deduced as
\begin{equation}
T^{'}_{H}=\frac{\Delta,_{r}(r_{+})}{4\pi
r^{2}_{+}(1+\tilde{\beta}\Sigma)}=T_{H}(1-\tilde{\beta}\Sigma),\label{n2}
\end{equation}
where
\begin{eqnarray*}
T_{H}&=&\frac{1}{r_+^3(r_+^4+8M_{pl}^2\beta_4)^2}\left[-16\mu
M_{pl}^2r_{+}\left(-3(\mu-1)\mu M_{pl}^2\beta_3^2\right.\right.\\
&+&\left.2(1-2\mu)\beta_4\right)(r-r_+)^2
(r_+^4+8M_{pl}^2\beta_4)+32M_{pl}^2\beta_4r_+^4(1-2\mu)(r-r_+)\\
&+&4\mu M_{pl}^2r_+\left(-3(\mu-1)\mu M_{pl}^2\beta_3^2
+2(1-2\mu)\beta_4\right)+r_+(\mu-1)(r-r_+)\\
&\times&(r_+^4+8M_{pl}^2\beta_4)^2+2r_+^2(r_+^4+8M_{pl}^2\beta_4)(1-2\mu)+4\mu M_{pl}^2(r-r_+)\\
&\times&\left.\left(-3(\mu-1)\mu M_{pl}^2\beta_3^2+2(1-2\mu)\beta_4\right)\right].
\end{eqnarray*}
which is the original Hawking temperature of the corresponding BH.
\section{Conclusion}
In this work, we have analyzed the tunneling probability and corrected
temperature $T^{'}_{H}$ of hairy BH for cubic and quartic interactions.
For this purpose, we utilized Hamilton-Jacobi ansatz and WKB
approximation and considered the modified Klein-Gordan
and Dirac equations for scalar and fermion particles, respectively. We
calculated corrected Hawking temperature for
scalar and fermion particles and
radiated temperature looked over preserved energy and charge. By
utilizing modified Klein-Gordan and Dirac equations, the corrected
Hawking temperature $T^{'}_{H}=T_{H}(1-\breve{\beta}\Xi)$ and
$T^{'}_{H}=T_{H}(1-\breve{\beta}\Sigma)$ vales have been
computed given in Eqs.(\ref{B8}) and (\ref{m8}) by considering the
effects of quantum gravity. We have concluded that the quantum
gravity effects increased the Hawking temperature. We
demonstrated that the corrected Hawking temperatures does not just
based on the properties of the BH, yet in addition depend upon the
quantum numbers, i.e., angular momentum, mass and energy of the transmitted
particles. The expression of corrected Hawking temperature in Eqs.(\ref{B8})
and (\ref{B9}) for scalar particle look similar to the expressioin in Eq.(\ref{m8})
and (\ref{m2}) for fermion particles but the angular momentum and mass are not same. Moreover, the black holes with the quartic interactions have more Hawking temperature than  BH's with cubic  interactions. We can conclude that the increasing number of interactions provide more radiation.

\acknowledgments

A. \"{O}.~acknowledges
financial support provided under the Chilean FONDECYT Grant No. 3170035. The authors thank an anonymous referees for their useful suggestions that helped improve the paper.

\end{document}